\documentstyle[12pt]{article}
\def\bh{\overline{h}}
\def\bd{\overline{d}}
\begin{document}

\title {\bf  GENERALIZED SCALING IN FULLY DEVELOPED  TURBULENCE}

\bigskip

\author{R. Benzi$^1$, L. Biferale$^1$,  S.Ciliberto$^2$, \\
M.V. Struglia$^1$,R. Tripiccione$^3$}
\maketitle

\bigskip
\bigskip
\centerline{$^1$ Dipartimento di Fisica, Universit\'a "Tor Vergata"}
\centerline{ Via della Ricerca Scientifica 1, I-00133 Roma, Italy}

\bigskip
\bigskip
\centerline {$^2$ Ecole Normale Sup\'erieure de Lyon
Laboratoire de Physique , C.N.R.S. URA1325}
\centerline {46 All\'ee d'Italie, 69364 Lyon, France }

\bigskip
\bigskip
\centerline{$^3$ INFN, Sezione di Pisa, S. Piero a Grado, 50100 Pisa, Italy}
\medskip
\bigskip
\bigskip
{\bf Abstract:}

In this paper we report numerical and experimental
results on the scaling properties of the velocity turbulent fields in
several flows.
The limits
of a new form of scaling, named Extended Self Similarity(ESS),
are discussed.
We show that, when a  mean shear is absent, the self scaling exponents
are universal and they do not depend on the specific flow (3D homogeneous
turbulence, thermal convection , MHD). In contrast, ESS is not observed
when a strong shear is present. We propose a generalized version
of self scaling which extends  down to the smallest
resolvable scales even  in cases where ESS is not present.
This new scaling is checked in several laboratory and numerical
experiment.
A possible  theoretical interpretation is also
proposed. A synthetic turbulent signal having most of the properties of a
real one has been generated.

\bigskip
\centerline{\it Submitted for publication in Physica D September 1995}

\newpage
\section{ Introduction }

In order to characterize the statistical properties
of fully developed turbulence\cite{Monin}, one usually studies the scaling
properties of moments of  velocity differences at  the scale $r$:
\begin{equation}
S_p(r)=<|v(x+r)-v(x)|^p>= <|\delta v(r)|^p>
\label{1.1}
\end{equation}
 where $<\cdots>$ stands for ensemble  average
  and $v$ is the velocity component parallel to
   $r$. At high Reynolds number $Re=U_0 L / \nu $
  the $S_p(r)$ satisfies the relation
\begin{equation}
S_p(r) \propto r^{\zeta(p)}
\label{1.2}
\end{equation}
 for $L>r >> \eta_k$ where $L$ is the integral scale, $\eta_k=(\nu^3/
 \epsilon)^{1/4}$ is the dissipative (Kolmogorov) scale, $\epsilon$ is the
mean energy dissipation rate, $\nu$ the kinematic viscosity and $U_0$
 the R.M.S. velocity of the flow.  The range of length
 $L>r >> \eta_k$, where  the scaling relation (\ref{1.2})
  is observed, is called the inertial range.
 The Kolmogorov (K41) theory \cite{K41} predicts $\zeta(p)=p/3$,
 but
 experimental\cite{Anselmet} and numerical \cite{Meneguzzi}
 results show that  $\zeta(p)$
 deviates  substantially from the linear law.
 This phenomenon is believed to be produced by  the intermittent
 behaviour of the energy dissipation \cite{K62} which can be taken into
account  by rewriting eq.(\ref{1.2}) in the following way:
\begin{equation}
 S_p(r) \propto  < \epsilon_r^{p/3} >  r^{p/3}
 \propto r^{\tau(p/3)+p/3}
\label{1.3}
\end{equation}
 where $\epsilon_r$ is the average of
 the local energy dissipation $\epsilon (x)$
   on a volume of size $r$ centered on a
   point $x$. A comparison of eq.(\ref{1.1}) and eq.(\ref{1.3}) leads to
   the conclusion that the scaling
   exponents $\tau(p/3)$ of the energy dissipation
   are related to those of $S_p$ by
   $\zeta(p)=\tau(p/3)+p/3$.

  Since the Kolmogorov (K62)
  theory  \cite{K62} many other models,  \cite{Frish}, \cite{Parisi},
\cite{Paladin}, \cite{Menev}
  \cite{Gagne}, \cite{She 1994} have been suggested
  to describe the behaviour of the $\zeta (p)$.
  However, it turns out that the $\zeta(p)$ may be not
  universal in non homogeneous, anisotropic flow and may depend on the
  location where measurements are done. Specifically, they may
  have  different values if one measures   either
  far away from boundaries, where turbulence is almost
  homogeneous and isotropic, or
  in locations of the flow where a strong mean shear is present.
  The $\zeta(p)$ depend also on the way in which
  turbulence is produced, for example  3D homogeneous turbulence,
  boundary layer turbulence, thermal convection and MHD.
  Thus there is the fundamental question of understanding
   in which way  all these parameters influence the scaling laws.
   Furthermore all the above mentioned models assume
   the existence of two well defined intervals of lengths
   that are the inertial range and a dissipation range.
   According to idea of multiscaling
   these two ranges
  may  eventually be connected by an intermediate region
  where the viscosity begins to act  \cite{Vergassola}.
   However this idea of a well defined inertial range, where
   viscosity does not act at all, and the idea of multiscaling
   turns out to be incompatible with the recently introduced new form
   of scaling, which has been named Extended Self Similarity (ESS)
    \cite{Benzi 1993A} \cite{Benzi 1995A}(see section 2 below).

  ESS has been observed
  in 3D homogeneous and isotropic turbulence
  both at low and high Re and  for a wide range of scales $r$ with
  respect to scaling (\ref{1.2}). In contrast
   ESS is not observed when a strong mean shear is present \cite{Stolov 1993}.
  All these experimental observations show that also the
  mechanisms by which
    energy is actually dissipated  in a flow are very poorly understood.
    Specifically one would like to understand how viscosity acts on different
    scales.
    This is clearly an important point in order to safely  use
    large eddy simulations in  real  applications.

  The purpose of this paper is to rationalize all the above mentioned
  results on scaling both in presence and absence of a shear.
   We propose a generalized form of ESS which has been checked
  in many different flows.  We have also  generated a signal
  which has all the statistical properties of a real turbulent signal.
  Our interpretation of ESS and of this  generalized scaling suggest that
  there is no  sharp viscous cut-off in the intermittent transfer
  of energy.

  The paper is organized as follows: in section 2 we remind
  the properties of ESS, in section 3 we discuss the  systems where
  ESS is not observed, in section 4 the hierarchy of structure
  functions is described, in section  5 the generalized form
  of scaling is discussed, in section  6 a possible theoretical
  interpretation  is proposed, in section 7 we discuss the multiscaling.
Finally conclusions are given in section 8.

\bigskip

\section{Extended Self Similarity }

Extended Self Similarity (ESS) is a property of velocity structure functions
of homogeneous and isotropic turbulence  \cite{Benzi 1993A,Benzi 1995A}.
It has been
shown using experimental and numerical data \cite{Benzi 1994}
that the structure functions present
an extended scaling range when one plots one structure function against the
other, namely:
\begin{equation}
  S_n(r) \propto S_m(r)^{\beta (n,m)}
\label{2.1}
\end{equation}
where $\beta (n,m) = \zeta (n)/\zeta (m)$.
The details  of ESS have been reported elsewhere \cite{Benzi 1995A}.
In the following we describe only the main features.

As an example  we consider two experimental data sets
at different $R_\lambda$,  which is
 the Reynolds number based
on the Taylor scale ($R_\lambda \simeq 1.4 \ Re^{1/2}$) \cite{Monin}.
The two experiments are
a jet at $R_\lambda=800$ and the wake behind a cylinder
at $R_\lambda=140$. In both cases data have been recorded  at about
25 integral scales downstream \cite{Benzi 1995A}.
In fig.1a
$S_6 / r^{\zeta(6)}$, computed for the two experiments, is
plotted as a function of $r$.
 In fig.1b  we show $S_3 / r^{\zeta(3)}$ as a function of $r$. In both
figures a  scaling region is observed only for the highest $R_\lambda$.
In contrast if the relative scaling (\ref{2.1}) is used, see fig.2,
a clear  scaling is  present for both $R_\lambda$ with $\beta(6,3)\cong 1.78$.
The vertical dashed lines in the fig.2 correspond to $r=5 \eta_k$
and they roughly indicate the extension of the scaling (\ref{2.1}), that
is $5 \eta_k < r < L$.

The ESS scaling has been checked both on numerical data
and in experiments, in a range $30 < R_\lambda < 2000 $.
A direct consequence of the scaling (\ref{2.1}) is  that for all p, $S_p$ can
be written in the following way:
\begin{equation}
 S_p(r)= C_p \  U_o ^p  \left [ {r \over L} f\left({r \over \eta_k }\right)
\right]^{\zeta (p)}
\label{2.2}
\end{equation}
with $U_o^3=S_3(L)$, $L=U_o^3/\epsilon$ being the integral scale and
$C_p$ dimensionless constants selected in such a way
that $f(x)=1$ for $x>>1$.
Eq.(\ref{2.2}) has been carefully checked by computing the function
$$f_p\left({r \over \eta_k }\right)  = \ { L\over r }  \
\left( {  S_p(r) \over C_p \  U_o ^p} \right) ^{1 \over  \zeta(p)}
$$
If $f_p$ is independent of p this means that eq.(\ref{2.2}) is satisfied.\\
This is seen in fig.3 where $\log( f_6 /f_2) $ and $\log( f_4 /f_2) $
are plotted as function
of $r /  \eta_k $. We clearly see that the both  ratios are  close
to 1 within $ 2\%$ for $r>5 \eta_k$.  This result shows that eq.(\ref{2.1})
is satisfied for  $5 \eta_k < r < L$.

ESS has been also checked for the temperature and velocity
fields in Rayleigh-Benard convection \cite{Benzi 1994B} and in the case
of a passive scalar\cite{Ruiz 1995}.
It turns out that ESS is a very useful tool in order to distinguish
between Kolmogorov and Bolgiano scaling \cite{Benzi 1994B}, \cite{Cioni 1995}.
In the case of the Bolgiano scaling it has been found that $\zeta(3)=2.08$
which is clearly very different from the Kolmogorov value $\zeta(3)=1$.
In spite of this large difference between the values of the
exponent  $\zeta(3)$, using ESS one discovers that the ratio
$\zeta(p)/\zeta(3)$
in the case of Bolgiano are equal to those of homogeneous and isotropic
turbulence. The same property   is observed for the $\zeta(p)$
obtained from measurements done on the solar wind \cite{Physics 1994} for MHD.
In Table I we compare the $\zeta(p)$ measured in different physical systems
and the ratios $ \beta (p,3)$ for MHD and Rayleigh-Benard convection.

Another interesting observation concerns the behaviour of $ \beta (p,3)$ with
respect to $ R_\lambda$. The values of $ \beta (p,3)$ reported in Table 1 have
been measured in the range $30<R_\lambda< 5 \ 10^6$. First we note that,
within error bars, any change or trend of $ \beta (p,3)$ as a function of
$ R_\lambda$ is absent.
Second, we show in fig.4 the dependence of $ \beta (6,3)$ on
$R_\lambda$ (a similar result has been reported in ref. \cite{Tab1995}).
This means that far away from boundaries the $\beta(p,3)$ are  constants
which do not depend on Re and on the way in which turbulence has been
generated.

A final point regarding ESS, concerns the generalization of the Refined
Kolmogorov Similarity Hypothesis.\\
The RKS hypothesis states that $ \epsilon_r \sim \delta v^3 / r $, as far as
concern the dependence on the scale $ r$, and supports eq.(\ref{1.3}).
We can generalize the RKS hypothesis by introducing an effective scale $L(r)=
S_3(r) / \epsilon $, as suggested by ESS, and we obtain the following relation:
$\epsilon_r = \delta v^3  \ \epsilon / S_3$.\\
Generalization of RKSH simply states that:
\begin{equation}
  S_p(r)=<|\delta  v(r)^p | > = {<\epsilon_r ^{p/3}> \over \epsilon^{p/3}} \
S_3(r)^{p/3}.
\label{2.3}
\end{equation}
In section 6 we give some theoretical support of eq.(\ref{2.3}).\\
Eq.(\ref{2.3}) has been first proposed in ref.\cite{Benzi 1995A} and carefully
checked in ref. \cite{Ruiz 1994}. A typical experimental
result is shown in fig.5 where
$<\epsilon^2>\  S_3^2$
 is plotted as a function of $S_6(r)$.
 The energy dissipation has been computed using the
 1-dimensional surrogate that is:
\begin{equation}
 \epsilon_r= {\nu \over r} \ \int_x^{x+r}
 \left({\partial V(x') \over \partial x}\right)^2  dx'
\label{2.4}
\end{equation}
In fig.5 one can
see a clear scaling extending  over almost ten decades from  the integral
scale to $\eta_k$. The slope of the straight
line is 1.005 showing that eq.(\ref{2.3}) is compatible with experimental
data.

One can argue that eq.(\ref{2.3}) is a trivial one because for
$r<\eta_k$, $\epsilon_r$ is constant and $S_p \propto r^p$, thus the scaling
$S_n \propto \ S_3^{p/3}$ is obviously satisfied.
Furthermore for $r$ in the inertial range eq.(\ref{2.3}) is certainly verified
because $(S_3/ \epsilon) \propto r$. However in principle
the proportionality
constant of eq.(\ref{2.3}) in the inertial and in the dissipative range could
be different. The fact that experimentally they are found equal
has several important consequences which will be discussed in
section 5.

\bigskip

\section{Systems where ESS is not observed}

In the previous section we have discussed several systems where
not only the ESS works  but also the exponents $\beta(n,3)$
are universal because they
do not depend on the systems and on $Re$.
We want to stress that this kind of universality, observed in different
flows, disappears if the system is influenced by  the presence
of a  strong mean  shear.  In this case ESS does not work,
because an extended range of scaling is not present  when $S_n$
is drawn as a function  of $S_3$.\\
Violation of ESS  has been observed experimentally  in boundary
layer turbulence \cite{Stolov 1993} \cite{Ruiz 1995A} and in the shear
behind a cylinder \cite{Benzi 1993B}.\\
In a recent numerical simulation \cite{Struglia 1995} the effect
of the shear on scaling laws  has been carefully investigated using a
Kolmogorov flow.

This simulation concerns a 3D fluid occupying a volume of $ V=L^3$ sites with
$L=160$, and forced such that the stationary solution
has a non-zero spatial dependent mean velocity
 $<\ v(\ x) >=
\hat x \sin ( {8 \pi z \over L}) $, where $\hat x$ is the versor in the
direction x, and L is the integral scale.\\
In figures 6a) and 6b) we show the standard ESS analysis by plotting
$\log (S_6)$ as a function of $\log (S_3)$ for two specific levels $z_a$ and
$z_b$, where $z_a$ and $z_b$  were chosen at the level of minimum and maximum
shear respectively.  The $R_\lambda$ of the simulation was 40 and
no scaling laws were present if examined as a function of  $r$.\\
Nevertheless, it is clear from figure 6a) that ESS is observed for the case
of minimum shear  and
it is not observed for the case of maximum shear (figure
6b). In both figures, the dashed lines are the best fit done in the range
between the $ 20$-th and $ 30$-th grid point and correspond to the slopes
$ \beta(6,3)=1.78$ and $ \beta(6,3)=1.43$ for the minimum and maximum shear
respectively.

However one finds that generalized Kolmogorov similarity hypothesis
eq.(\ref{2.3}) is satisfied also for values of $r$ where ESS is no longer
satisfied. In order to highlight the previous comment we consider  again
the above mentioned Kolmogorov flow. In fig.7a and 7b we show the result
of the scaling obtained by using eq.(\ref{2.3}) at the correspondent z-levels
of figures 6a and 6b for p=6. As one can see the generalized Kolmogorov
similarity hypothesis is well satisfied in both cases although for $z_b$ ESS
is not observed.
This is another important experimental and numerical result
which we will consider again  in  section 5.

The other relationship which we have observed to hold from large to small
scales even in absence of ESS is the moment hierarchy recently proposed
in ref.\cite{She 1994} and rewritten in terms of velocity structure functions
in ref.\cite{Ruiz 1995B}.

\bigskip

\section{Hierarchy of structure functions }

In  a  recent letter \cite{She 1994} She  and Leveque have proposed
an interesting theory
to explain the anomalous scaling exponents of velocity structure
functions.
The theory yields a prediction
$$\zeta(p)=p/9+2 ( 1 - (2/3) ^{p/3})$$
which is in very good agreement with  available experimental
data\cite{Benzi 1995A}.

The She Leveque model is based upon
the  fundamental
assumption on the hierarchy of   the
moments, $<\epsilon_r ^{n} >$,
of the local energy dissipation.
Specifically they consider that:
\begin{equation}
 {<\epsilon_r ^{n+1} > \over <\epsilon_r ^{n} >}=
A_n \  \left( {<\epsilon_r ^{n} > \over <\epsilon_r ^{n-1} >}  \right) ^\beta
\ \big( \epsilon_r ^{ ( \infty ) } \big)  ^{(1 - \beta)}
\label{4.1}
\end{equation}
where $A_n$ are geometrical constants and
$ \epsilon_r ^{ ( \infty ) } =\lim_{n\rightarrow \infty }
({<\epsilon_r ^{n+1} > \over <\epsilon_r ^{n} >})$ is
associated in ref.\cite{She 1994} with  filamentary structures of the flow.
On the  basis of simple arguments it is assumed that:
$\epsilon_r ^{ ( \infty ) } \propto r^{-2/3}$.
The value of $\beta$ predicted in ref.\cite{She 1994} is $2/3$.
Notice that in eq.(\ref{4.1}) for n=1,
taking into account that $<\epsilon_r>= \epsilon $
is  constant  in $r$, one immediately finds that
\begin{equation}
\big( \epsilon_r ^{ ( \infty ) } \big)  ^{(1 - \beta)} \propto
  <\epsilon_r^2> ={ S_6 \over S_3^2}
\label{4.2}
\end{equation}
  where eq.(\ref{2.3}) has been used.

Equation (\ref{4.1}), which  has been experimentally tested in ref.\cite{Ruiz
1995C},
can be extended to the velocity structure functions\cite{Ruiz 1995B}.
Taking in eq.(\ref{4.1}) the value $n=p/3$ and using equations (\ref{2.3})
and (\ref{4.2}) after some algebra one finds  the following relation for
the  velocity structure functions:
\begin{equation}
  F_{p+1}(r) = C_p  \ \big( { F_{p}(r) }
\big) ^{\beta '}  \cdot  \tilde F(r)
\label{4.3}
\end{equation}
where $$ F_{p+1}(r) = { S_{p+1}(r) \over S_p (r) }$$ and $$
\tilde F(r)=
\left( { S_6 \over S_3^{(1+\beta)} } \right) ^{(1 - \beta ')
\cdot  \over 3 (1-\beta)}, $$
$C_p$  are geometry dependent constants and
$\beta '=\beta ^{1/3}$.

Notice that eq.(\ref{4.3}) is certainly valid for any $\beta$
in the dissipative range where $S_n \propto r^n$.
Equation (\ref{4.3}) has been experimentally tested in  ref.\cite{Ruiz 1995B}.

This can be seen in figures 8a) and 8b) where the scaling obtained
for various  p using eq.(\ref{4.3}) is reported for two different $Re$.
As we have already observed in the case of eq.(\ref{2.3})(figure 5),
the scaling extends from large to small scales even
for values of $r$ where ESS is no longer satisfied.

\bigskip

\section{A generalized form of ESS }

In sections  3) and 4) we have shown that the GKRS  eq.(\ref{2.3}) and
the  hierarchy of moments eq.(\ref{4.3})
are two relations which are satisfied even in flows where
ESS is not observed.  These results suggest that the concept of
ESS could be generalized in such a way to take into account the
scaling relations equations (\ref{2.3}) and (\ref{4.3}) properly.

For this purpose we introduce the dimensionless structure function
\begin{equation}
 G_p(r)= { S_p(r) \over S_3(r)^{p/3}}
\label{5.1}
\end{equation}
According to Kolmogorov theory (\ref{5.1}) should be a constant both in the
inertial and in the dissipative range, although the two constant are
not necessarily the same. Because of the presence of anomalous
scaling $G_p(r)$ are no longer constants
and by using (\ref{2.3}) we have:
\begin{equation}
 G_p(r) = < \epsilon_r ^{p/3} >
\label{5.2}
\end{equation}
Thus  the functions $G_p(r)$ satisfy the hierarchies (\ref{4.1}) and
(\ref{4.3}).
Following the results of sections 3) and 4) equation (\ref{5.2}) is valid
for all scales even in cases where ESS is not  verified. Therefore,
it seems reasonable to study the  self scaling properties of
$G_p(r)$ or, equivalently, the self-scaling properties of the energy
dissipation averaged on an interval of size $r$:
\begin{equation}
 G_p(r) = G_q(r)^{\rho(p,q)}
\label{5.3}
\end{equation}
where we have by definition:
\begin{equation}
 \rho(p,q) = { \zeta(p) - p/3 \ \zeta(3) \over \zeta(q) - q/3 \ \zeta(3)}
\label{5.4}
\end{equation}
$ \rho(p,q)$ is given by the ratio between deviation from the K41 scaling.
It will play an essential role in our understanding of energy cascade.
Indeed, it is easy to realise that it is the only quantity that can
stay constant along all the cascade process: from the integral to
the sub-viscous scales.  It is reasonable  to imagine that the
velocity field becomes laminar in the sub-viscous range,
$S_p(r) \propto r^p $, still preserving some intermittent degree
parametrized by the ratio between corrections to K41 theory.
In order to  check the validity of eq.(\ref{5.3}) we have plotted in fig.9
$G_6(r)$ versus $G_5(r)$ for many different  experimental set-up
\cite{Benzi 1993B},\cite{Ruiz 1995A}, \cite{Benzi  1994B}, \cite{Benzi 1995B},
done at different
Reynolds numbers and for some direct numerical simulation with
and without large scale shear.
As one can see the straight line behaviour is very well
supported within experimental errors (of the order of $3\%$) no
deviations from the scaling regime are detected. Similar results
are obtained, using different $G_p(r)$ and $G_q(r)$.

There are two alternative ways to check (\ref{5.3}).
First of all, one can rewrite it
in the following way:
\begin{equation}
S_p(r) = (S_q(r))^{\rho(p,q)} (S_3 (r))^{p/3 - r(p,q) q/3}
\label{5.5}
\end{equation}
If (\ref{5.3}) is true then $ \rho(p,q) $ should be equal to
$ r(p,q) $. One can use (\ref{5.5}) directly and perform a two variable
fit of $ \rho (p,q) $ and $ r(p,q)$. Then the quantity:
\begin{equation}
\sigma_{p,q} = \frac{\rho(p,q) - r(p,q)}{\rho(p,q)}
\end{equation}
gives a measure of the accuracy of (\ref{5.3}). We have computed
$\sigma_{p,q}$ for $p$ and $q$ in the range [1,8] for all the experimental
and numerical results. We have found that:
\begin{equation}
\sup_{p,q} [ \sigma_{p,q} ] = 0.01
\end{equation}
Where the above test has been done over all the experimental and numerical
data available to us.
This result tells us that the accuracy of (\ref{5.3}) is extremely well
verified.

A second and independent check of (\ref{5.3}) can be done by using
(\ref{4.3}). Indeed (\ref{4.3}) can be checked either for fixed $p$ as a
a function of $r$ or for fixed $r$ as a function of $p$. In the second case
we may assume that the constant $C_p$ in (\ref{4.3}) is p-independent and,
by plotting in a log-log scale $F_{p+1}$ against $F_p$ for fixed $r$
and different
$p$, we can estimate the exponent $\beta '$. If (\ref{5.3}) is true
than we should observe scaling (\ref{4.3}) both at large scale and
very small scale with the same value of $\beta '$. Let us remark  that
the previous statement (on which our test of (\ref{5.3}) is based)
depends on the two assumptions that the log-poisson hierarchy for
structure functions is true and that the constant $C_p$ in (\ref{4.3})
are p-independent.
In fig. 10a and 10b we show a log-log
plot of $F_{p+1}$ against $F_p$ for $p=1,...6$ and
$r= 3 \eta_k $ and $r= 30 \eta_k$ for
the case of two numerical simulations, namely Rayleigh Bernard thermal
convection and channel flow. As one can see a clear scaling is
observed with the same scaling exponent $\beta '$ both for small and
relatively large values of $r$. This confirms the quality of the
generalized ESS scaling (\ref{5.3}).

\bigskip

\section{A theoretical interpretation }

The aim of this section is to discuss a possible theoretical interpretation of
the experimental and numerical results previously shown. Our starting point is
to revise the concept of scaling in fully developed turbulence.

Let us consider three length scales $ r_1> r_2> r_3$ and our basic
variables to describe the statistical properties of turbulence, namely the
velocity difference $ \delta v(r_i)$. We shall restrict ourself to those
statistical models of turbulence based on random multiplier.\\
Thus we shall assume that there exists a statistical equivalence of the form:
\begin{equation}
\delta v(r_i) = a_{ij} \delta v(r_j)
\label{6.1}
\end{equation}
where $ r_i< r_j$ and $ a_{ij}$ is a random number with a prescribed
probability distribution $ P_{ij}$.\\
By definition, we have:
\begin{equation}
 a_{13} = a_{12} a_{23}
\label{6.2}
\end{equation}
Equation (\ref{6.2}) is true no matter which is the ratio $\frac{r_1}{r_2}$
and $\frac{r_2}{r_3}$. Now we ask ourselves the following question: what is
the probability distribution $P_{ij}$ which is functionally invariant under
the transformation (\ref{6.2})? This question can be answered by noting that
equation (\ref{6.2}) is equivalent to write:
\begin{equation}
 \log a_{13} = \log a_{12} + \log a_{23}
\label{6.3}
\end{equation}
(we assume $a_{ij}> 0$). Thus our question is equivalent to ask what are the
probability distribution stable under convolution. For independently
distributed random variables a solution of this problem can be given in a
complete form \cite{Feller}, \cite{Novikov 1994}. If the variables are
correlated the situation becomes much more difficult to solve, as it is well
known from the theory of critical phenomena. For the time being we shall
restrict ourselves to independent random variables.

In this case, for instance, the gaussian and the Poisson distributions
are well known
examples of probability distribution stable under convolution. These two
examples correspond to two
turbulence model proposed in literature, namely the log-normal model
\cite{K62} and the log-Poisson model \cite{She 1994}, \cite{Dubrulle 1994},
\cite{She Waym}, \cite{Cast }.
A more general description can be found in \cite{Novikov 1994}.

We can have a different point of view on our question which  is fully
equivalent to the above discussion. A simple solution to our question is given
by all probability distribution $P_{ij}$ such that:
\begin{equation}
\langle a_{ij}^p \rangle \equiv \prod_{k=1}^n \left( \frac{g_k(r_j)}{g_k(r_i)}
\right)^{\gamma_k(p)}
\label{6.4}
\end{equation}
for any functions $g_k(r_i)$ and $\gamma_k(p)$ ($\langle \cdots \rangle$
represents average over $P_{ij}$). Indeed we have:
\begin{equation}
 \langle a_{13}^p \rangle = \langle a_{12}^p a_{23}^p\rangle = \prod_{k=1}^n
 \left( \frac{g_k(r_1)}{g_k(r_2)}\right)^{\gamma_k(p)}\prod_{k=1}^n
 \left( \frac{g_k(r_2)}{g_k(r_3)}\right)^{\gamma_k(p)}= \prod_{k=1}^n
 \left( \frac{g_k(r_1)}{g_k(r_3)}\right)^{\gamma_k(p)}
\label{6.5}
\end{equation}
We want to remark that equation (\ref{6.4}) represents the most general
solution to our problem, independent of the scale ratio $r_i/r_j$.\\
Let us give a simple example in order to link equation (\ref{6.4}) to the case
of probability distribution stable under convolution. Following \cite{She
1994}, \cite{Dubrulle 1994} and \cite{She Waym} let us consider the case of a
random log-Poisson  multiplicative process, namely:
 \begin{equation}
  a_{ij}=  A_{ij} \beta ^x
\label{6.6}
\end{equation}
where $x$ is a Poisson process $P(x=N)=\frac{C_{ij}^N e^{-C_{ij}}}{N!}$.\\
By using (\ref{6.6}) we obtain:
\begin{equation}
 \langle a_{ij}^p \rangle = A_{ij}^{p} \exp (C_{ij}(\beta ^p -1))
\label{6.7}
\end{equation}
Equation (\ref{6.7}) is precisely of the form (\ref{6.4}) if we write
\begin{equation}
 A_{ij}=\frac{g_1(r_j)}{g_1(r_i)}~~~~~~~\exp C_{ij}=\frac{g_2(r_i)}{g_2(r_j)}
\label{6.8}
\end{equation}
In order to recover the standard form of She-Leveque model we need to assume
that (see also \ref{6.14}):
\begin{equation}
  g_1(r_i)\sim {r_i}^h
\label{6.9}
\end{equation}
\begin{equation}
 g_2(r_i)\sim r_i^{2}
\label{6.10}
\end{equation}
This example highlights one important point in our discussion, i.e. the
general requirement of scale invariant random multiplier (\ref{6.4}) does not
necessary imply a simple power law scaling as expressed by the equations
(\ref{6.9}-\ref{6.10}). Moreover, the general expression (\ref{6.4}) is
compatible only to infinitively divisible distribution. For instance, previous
random multiplier model for turbulence, such as the $\beta$-random model or
the p-model, cannot be expressed in the general form (\ref{6.4}) independently
of the ratio $r_i/r_j$.

It is worthwhile to review the multifractal language at light of the previous
discussion. In the multifractal language for turbulence, the two basic
assumptions are:

I) The velocity difference on scale $r$ shows local scaling law with exponent
$h$, i.e. $\delta v(r)\sim r^h$;

II) the probability distribution to observe the scaling $\delta v(r)\sim r^h$
is given by $r^{3-D(h)}$.

In the multifractal language, therefore, there are two major ansatz: one
concerns power law scaling of the velocity difference (assumption I) and the
other one concerns a geometrical interpretation (the fractal dimension $D(h)$)
of the probability distribution to observe a local scaling with exponent $h$.
How is it possible to generalize the multifractal language in order to take
into account equation (\ref{6.4})?

 As we shall see, the theory of infinitively divisible
distribution is the tool we need to answer the previous question.
All published model of turbulence based on infinitively divisible distribution
are equivalent to write $D(h)$ in the form:
\begin{equation}
 3-D(h)=d_0 f\left[\frac{h-h_0}{d_0}\right]
\label{6.11}
\end{equation}
where $d_0$ and $h_0$ are two free parameters while the function $f(x)$
depends only on the choice of the probability distribution. For instance for
log-normal distribution $f(x)=x^2$. Equation (\ref{6.11}) allows us to write:
\begin{equation}
\langle\delta v(r)^p\rangle=\int d\mu (h) r^{hp}r^{3-D(h)} = r^{h_0p+d_0 H(p)}
\label{6.12}
\end{equation}
where
\begin{equation}
  H(p)= \inf_x(px+f(x))
\label{6.13}
\end{equation}
We can see that equation (\ref{6.12}) is equivalent to a random
multiplicative process given by:
\begin{equation}
 \langle a_{ij}^p\rangle=\left(\frac{r_j}{r_i}\right)^{h_0p}\left(\frac{r_j}
{r_i} \right)^{d_0H(p)}
\label{6.14}
\end{equation}
Equation (\ref{6.14}) can be generalized to the form (\ref{6.4}) by allowing
$h_0$ and $d_0$ to depend on $r$, i.e.
\begin{equation}
 \langle a_{ij}^p\rangle=\frac{\left(r_j^{\bh_0(r_j)}\right)^p}{\left(
r_i^{\bh_0(r_i)} \right)^p}\left[\frac{r_j^{\bd_0(r_j)}}{r_i^{\bd_0(r_j)}}
\right]^{H(p)}
\label{6.15}
\end{equation}
where:
\begin{equation}
\bh_0(r) = h_0 s_h (r) ~~~~~~~ \bd_0(r) = d_0 s_d(r)
\label{6.15bis)}
\end{equation}
Equation (\ref{6.15})
is equivalent to (\ref{6.4}) by using:
\begin{equation}
 g_1(r_i) = r_i^{\bh_0(r_i)} ~~~~~~~ g_2(r_i)=r_i^{\bd_0(r_i)}
 \label{6.16}
\end{equation}
\begin{equation}
 \gamma_1(p) =p
 \label{6.17}
\end{equation}
\begin{equation}
\gamma_2(p) = H(p)
\label{6.18}
\end{equation}
The same results can be obtained by (\ref{6.12}), i.e. we have
\begin{equation}
 \langle \delta v(r)^p\rangle=r^{\bh_0(r)p+\bd_0(r)H(p)}
 \label{6.19}
\end{equation}
Note that the saddle point evaluation of (\ref{6.12}) is not spoiled by the
dependence of $h_0$ and $d_0$ on $r$.

We have seen that (\ref{6.4}) can be reformulated in terms of multifractal
language for infinitively divisible distribution whose function $D(h)$ can be
rewritten as in (\ref{6.11}). We can ask the following question: what is the
physical meaning of (\ref{6.4}) or its multifractal analogous (\ref{6.15}-
\ref{6.19})? It is precisely the multifractal language which allows us to
answer this question.
Indeed, the two basic assumption for the multifractal language
can now be replaced in the following way:

I) the velocity difference on scale $r$ behaves as
\begin{equation}
\delta v(r)\sim g_1(r)
g_2(r)^x;
\label{eq:33.a}
\end{equation}

II) the probability distribution to observe I is $g_2(r)^{f(x)}$.\\
 Then we have
\begin{equation}
\langle \delta v(r)^p\rangle=\int d\mu (x) g_1(r)^p g_2(r)^{px+f(x)}=g_1(r)^p
 g_2(r)^{H(p)}
 \label{6.20}
\end{equation}
by employing a saddle point integration.
The most clear physical interpretation of (\ref{6.20}) is that the probability
to observe a given fluctuation of the velocity difference has no more
geometrical interpretation linked to the fractal dimension $D(h)$.
The probability distributions are controlled by a dynamical variable $g_2(r)$
which at this stage we still need to understand. An insight on the dynamical
meaning of $g_2(r)$ can be obtained by the following considerations.

Let us define $\epsilon (r)$ the average of the energy dissipation on a scale
$r$. We can define the eddy turnover time $\tau (r)$ on scale $r$ as:
\begin{equation}
 \frac{\delta v^2(r)}{\tau(r)}\sim \epsilon(r)
\label{6.21}
\end{equation}
We have seen that all experimental and numerical data suggest that the
following relation is always ( see also eq.6):
\begin{equation}
 \frac{\epsilon(r)}{\langle\epsilon\rangle}= ^s \frac{\delta v^3(r)}{\langle
 \delta v^3(r)\rangle}
\label{6.22}
\end{equation}
where $=^s$  means that all moments on the r.h.s. are equal to l.h.s. By using
(\ref{6.21}-\ref{6.22}) we obtain the definition of length $L(r)$:
\begin{equation}
 L(r) \equiv \delta v(r)\tau(r)=\frac{\langle \delta v^3(r)\rangle}{\epsilon}
\label{6.23}
\end{equation}
$L(r)$ cannot be regarded as a real length scale in the physical space. Rather,
$L(r)$ should be considered as a dynamical variable entering into the
statistical description of turbulence. This is precisely the idea behind ESS
which reformulate the scaling properties of turbulence in terms of $L(r)$.
Indeed in order to obtain ESS from (\ref{6.20}) it is sufficient to state
that, within the range of scales where ESS is observed, $g_1(r)^{1/ h_0}
\sim
g_2(r)^{1/d_0}\sim
L(r)$.
The physical meaning of ESS is strictly linked to (\ref{6.23}) and in
particular to (\ref{6.22}) which is a generalization of Kolmogorov Refined
Similarity Hypothesis.\\
Let us summarize all our previous findings:\\
A) we have introduced the idea of scale invariant random multiplier satisfying
equation (\ref{6.4});\\
B) we have shown that infinitively divisible distributions are all compatible
with(\ref{6.4});\\
C) we have shown that the multifractal language specialized for the case of
infinitively divisible distribution gives equation (\ref{6.4}) (with $n=2$ and
$\gamma_1(p)$ linear in $p$) and it is equivalent to scale invariant random
multiplier;\\
D) finally we have argued that the correct scaling parameter to describe the
statistical properties of small scale turbulent flows is not directly linked
to a simple geometrical interpretation, rather it should be considered a
dynamical variable.\\
Our finding  A-D enables us to have a unified theoretical interpretation of
the experimental and numerical results presented at the beginning of this
paper.
Indeed equation (\ref{6.19}) or (\ref{6.20}) tells us that the anomalous part
of the structure functions:
\begin{equation}
 G_p(r)\equiv \frac{\langle\delta v^p(r)\rangle}{\langle\delta v^3(r)\rangle^
 {p/3}}
\label{6.24}
\end{equation}
satisfies the scaling properties:
\begin{equation}
 G_p(r)= G_q(r)^{\rho_{p,q}}
\label{6.25}
\end{equation}
where $\rho_{p,q}\equiv \frac{\zeta_p -p/3}{\zeta_q-q/3}$. According to our
analysis of the experimental and numerical results, the scaling (\ref{6.25})
is observed down to the smallest resolved scale.\\
We have shown that, in the theoretical framework so far exposed, we recover
the ESS when $ g_1(r)^{1/h_0} \sim g_2^{1/d_0} \sim L(r)$.
If $ g_1(r)^{1/h_0}\neq g_2(r)^{1/d_0}$ we lose ESS, but
its generalized version (\ref{6.25}) is still valid.\\

\subsection{Synthetic Turbulence}

We can also use (\ref{6.19})and (\ref{6.20}) to simulate a synthetic signal
according to a random multiplicative process satisfying (\ref{6.15}). This can
be done by using the algorithm recently introduced in \cite{bbcpvv}.\\
Let us consider a wavelet decomposition of the  function $ \phi (x)$:
\begin{equation}
\phi(x)=\sum_{j,k=0}^{\infty} \alpha_{j,k} \psi_{j,k}(x)
\label{6.26}
\end{equation}
where $\psi_{j,k}(x)= 2^{j/2} \psi(2^jx -k)$ and $\psi(x)$ is any wavelet
with zero mean. The above decomposition defines the signal
as a diadic superposition of basic fluctuations with different
characteristic widths (controlled by the index $j$) and centered in
different spatial points (controlled by the index $k$).
 For functions defined on $N=2^n$ points in
 in the interval $[0,1]$
 the sums in (\ref{6.26}) are restricted from zero to $n-1$ for
 the index $j$ and from zero to $2^j-1$ for $k$ \cite{meyer}.

 In \cite{bbcpvv}
 it has been shown that the statistical behavior of signal increments:
  $$<|\delta \phi(r)|^p> =
 <|\phi(x+r)- \phi(x)|^p> \sim r^{\zeta(p)}$$
is controlled by the coefficients $\alpha_{j,k}$. By defining the
 $\alpha$ coefficients in terms of
  a multiplicative random process on the diadic tree
  it is possible to give an explicit expression for the scaling
  exponents $\zeta(p)$. For example,  it is possible to recover
  the standard anomalous scaling by defining the $\alpha$'s tree in term
  of the realizations of a random variable $\eta$ with a probability
   distribution
  $P(\eta)$:
$$ \alpha_{0,0}$$
  $$
\alpha_{1,0}  =   \, \eta_{1,0} \, \alpha_{0,0}; \,\,\,
\alpha_{1,1}  =  \, \eta_{1,1} \, \alpha_{0,0}; \nonumber $$
\begin{equation}
\alpha_{2,0} = \,\eta_{2,0} \,\alpha_{1,0}; \,\,
\alpha_{2,1} = \,\eta_{2,1} \,\alpha_{1,0}; \,\,
\alpha_{2,2} = \,\eta_{2,2} \,\alpha_{1,1}; \,\,
\alpha_{2,3}=  \,\eta_{2,3} \,\alpha_{1,1},
 \label{eq:alb}
\end{equation}
and so on. Let us note that in the previous multiplicative
process different scales are characterized by
different values of the index $j$, i.e. $r_j =2^{-j}$.
If the $\eta_{j,k}$ are {\it i.i.d.} random variable it is
straightforward to realise that $\alpha_{j,k}$ are random variables
with moments given by:
\begin{equation}
<|\alpha_{j,k}|^p> = r_j^{- \log_2(\overline{\eta^p})}
\label{eq:scal_alfa}
\end{equation}
where   the ``mother eddy' $\alpha_{0,0}$
 has been chosen equal to one. In (\ref{eq:scal_alfa})
 with $\overline{\cdots}$ we intend averaging over the $P(\eta)$
distribution.
In \cite{bbcpvv} it has been shown that also the signal $\phi(x)$
 has the same  anomalous  scaling of (\ref{eq:scal_alfa}).

In order to generalize this construction for function showing ESS or
generalized-ESS scaling of the form (\ref{6.19}) and (\ref{6.25})
is now sufficient to take a probability distribution, $P_l(\eta)$,
 for the random multiplier with the appropriate scale dependency
 (\ref{6.4}). This will be implemented by allowing a
 dependency of $P(\eta_{jk})$
 on the scale $r_j=2^{-j}$, i.e. the  $\eta$'s random variables will be
 still independently distributed but not identically distributed
 with respect to variation of the scaling index $j$.\\
According to the previous discussion, ESS corresponds to have only
one seed-function defining the multiplicative process, i.e.
$ g_1(r)^{1/d_0}\neq g_2(r)^{1/h_0}$
in the range of scales where ESS is valid ($r \le 5 \eta_k$).
On the other hand,
at scales smaller then $5 \sim 6 $ Kolmogorov scale, ESS is not more valid
because  $g_2^{1/d_0}$ begins to deviate substantially from
$g_1^{1/h_0}$: only G-ESS
should be observed and we need a multiplicative process defined in terms
of two different seed-functions.\\
Following this recipe we  define the  signal such that :
\begin{equation}
<\delta v(r)^p > = U_0^p F(r)^{p/3} G(r)^{\zeta(p)-p/3}
\label{eq:monstre}
\end{equation}
where
\begin{equation}
F(r) = <\delta v(r)^3>/U_0^3
\label{eq:FG}
\end{equation}
The function $G(r)$ is defined in such a way that for $F(r)$ much greater
than $\eta_k / L$, $G(r) \sim F(r)$ while for very small scales $r$
we have $G(r) \sim \eta_k/L$. In the following we choose
the simplest ansatz:
\begin{equation}
G(r) = B +A F(r)
\label{eq:FGG}
\end{equation}
with $B = \eta_k/L$ and $A$ is a dimensionless constant.

Let us now spend some words in order to clarify
the previous definitions.
Relation (\ref{eq:monstre}) is defined such that
experimental results are reproduced with good accuracy and
G-ESS scaling (\ref{6.25})
is satisfied by definition.
By assuming (\ref{eq:FGG}) the only unknown function is
$F(r) = <\delta v(r)^3>/U_0^3 $.
On the other hand the function $<\delta v(r)^3>$
 is always very well fitted
by the Batchelor parameterization:
\begin{equation}
 <\delta v(r)^3> = \frac{U_0^3}{L \eta_k^2}
 \frac{r^3}{(1 +(r/\eta_k)^2)}.
 \label{eq:dv3}
 \end{equation}

 {}From expression (\ref{eq:monstre}) is  immediate to
 extract the expression for the two seed-functions $g_1(r), g_2(r)$
 used in the previous sections, namely:
 \begin{eqnarray}
 g_1(r) &= \left(\frac{F(r)}{G(r)}\right)^{1/3} G(r)^{h_0} \nonumber \\
  g_2(r) &= G(r)^{d_0}
  \label{eq:g1g2}
  \end{eqnarray}
  Let us note that $g_1(r)$
  goes smoothly from the intermittent value, $g_1(r) \sim r^{h_0}$
($h_0= 1/9$ for the
  case of She Leveque model),  assumed
  in the inertial range to the
  laminar value, $g_1(r)=\sim r $,
characteristic of  scales much smaller than
  Kolmogorov scale.

 For the practical  point of view we have  constructed
 our signal by using  a random process for the multiplier $\eta_{j,k}(r_j)$
with a scale-dependent Log-Poisson distribution. The scale dependency of
parameters entering the distribution has been fixed in
terms of relations (\ref{eq:g1g2}) and (\ref{6.8}) and such that the
$\zeta(p)$ exponents correspond to the She-Leveque \cite{She 1994}
expression, namely:
\begin{equation}
\eta_{jk}(r_j) = A_{j,j+1} \beta^{x_{j,j+1}}
\label{eq:poisson}
\end{equation}
where $x_{j,j+1}$ is a Poisson variable with mean $C_{j,j+1} =
 \log(g_2(r_{j+1})/g_2(r_{j}))$, $A_{j,j+1} = g_1(r_j)/g_1(r_{j+1})$ and
$\beta=2/3$. This choice leads to the standard Log-Poisson scaling
in the inertial range:
$$\zeta(p) = h_0 p + (1-3h_0)\frac{(1-\beta^{p/3})}{(1-\beta)}$$
and to the following expression for the ratios of deviations
to the Kolmogorov law:
$$ \rho_{p,q}= \frac{H(p)-p/3}{H(q)-q/3}, \; \; \; H(x)=
\frac{1-\beta^{x/3}}{(1-\beta)} $$.

 Signal constructed according to this scenario will be referred to as signal-A
 in the following.\\
    In fig. 11 we show the structure function of order 6 for such a signal
  plotted  versus the separation scale $r$ at moderate Reynolds
  number. Clearly, for this choice of Reynolds number there is not any
  inertial range of scale where scaling exponents could be
  safely measured. On the other hands, our signal shows G-ESS scaling,
  how it is possible to see in fig. 12.

 \section{Multiscaling}

We now turn our attention to a different question which is connected to the
theoretical results so far discussed, namely the role played by viscous
effects.
It is generally argued that the anomalous scaling can be observed for scale
larger than a given viscous cutoff. The physical interpretation of this
statement is that non linear, intermittent, transfer of energy is acting only
for scale larger than the viscous scale. Below such a scale the structure
function are supposed to show a simple (regular) scaling $\langle\delta v^p(r)
\rangle\sim r^p $.

Usually the viscous cutoff is introduced as the scale at which the local
Reynolds number is of order one, namely:
\begin{equation}
\frac{\delta v(r) r}{\nu}\sim 1
\label{6.27}
\end{equation}
This condition can be obtained by the requirement that the local energy
transfer $ \epsilon(r)\sim (\delta v^3(r)/r)$ becomes equal to the energy
dissipation $ \nu (\delta v^2(r)/r^2)$:
\begin{equation}
 \nu \frac{\delta v^2(r)}{r^2} \sim \frac{\delta v^3(r)}{r}
\label{6.28}
\end{equation}
which gives equation (\ref{6.27}). There is a well defined prediction,
based on (\ref{6.27}), formulated by Frisch and Vergassola using the
multifractal language. Indeed for any exponent $ h$ one can introduce the $
h$-dependent viscous cutoff given by:
\begin{equation}
r_d^{h+1}\sim \nu
\label{6.29}
\end{equation}
where $ \delta v(r)\sim r^h$.\\
It follows that $ r_d(h)$ is a fluctuating quantity. There are two consequences
of this theory. The first one predicts that for the structure functions $
\langle \delta v^p(r)\rangle$ there exists a cutoff scale $ r_p$ dependent on $
p$ and moreover $ r_p<r_q$ for $ p>q$.\\
The second prediction concerns the moment of the velocity gradients $ \Gamma$
which are:
\begin{equation}
\langle \Gamma^p\rangle \sim \langle r_d(h)^{(h-1)p} r_d(h)^{3-d(h)}\rangle
\sim Re^{-z(p)}
\label{6.30}
\end{equation}
with $ z(p)= \sup_h \left[ \left( (h-1)p +3 -D(h)\right)/(1+h)\right]$.\\
Between the two predictions the first one is qualitatively more peculiar
of (\ref{6.29}).\\
In particular, the first prediction states that between the end of the
inertial range (i.e. the region where anomalous scaling of $ \langle \delta
v^p(r)\rangle$ with respect to $ r$ is detected) and the dissipation cutoff
$ r_d$, the local slope is controlled in rather complicate way by $ D(h)$.
The second prediction is somehow weaker because present experimental data do
not distinguish among several models, so far proposed, for the Re-dependence
of $\langle \Gamma^p\rangle$.

In order to compare the multiscaling in the dissipation range with our
experimental and numerical data, we have produced a synthetic turbulence
signal (signal B hereafter) similar to the one already discussed but with $
d_0$ and $ h_0$ independent on $ r$. The effect of dissipation is introduced
by using (\ref{6.27}). In fig.13 we compare the local scaling exponents $
d(\log\langle\delta v^p\rangle)/d(\log r)$ for $ p=6$ between the two
synthetic signals.\\
In fig.14 we plot the relation (\ref{6.22}) for $ p=6$ for signal A and B.
Finally in fig.15 we plot the ratio $G_4/G_6^{\rho_{4,6} }$ for signal B.
By comparing figures 13, 14 and 15 with the analogous experimental and
numerical results discussed in the previous sections (see fig 5,7,9), we can
state that the quantitative and qualitative prediction based on (\ref{6.29})
is not verified by experimental data. On the other hand, signal A, based on an
explicit $r$-dependence of $ d_0$ and $ h_0$, seems to be more closely related
to what observed experimentally. Let us remark that signal A has no cutoff
effect imposed by condition (\ref{6.27}).

The above discussion rules out the effect of multiscaling on the viscous
cutoff (\ref{6.29}).\\
Previous claims on the validity of multiscaling effects should be considered
either wrong or affected by experimental errors. On the other hand our model,
used to implement the synthetic signal A, should be considered a very accurate
model even for scale close to the regular region where $ \delta v(r)\sim r$.

There is, however, a theoretical question concerning multiscaling which we are
still not able to answer completely and that we shall try to formulate in the
following.\\
There are two possible scenario in which a viscous cutoff may be considered.

In the first scenario (let us call it scenario I) we can imagine to consider
equation (\ref{6.27}) as a fundamental relation independent of any other
theoretical considerations. The idea is that when the local Reynolds number is
sufficiently small, then non linear effect must be neglected. In order to
compute the viscous scale, one should make use of the relation:
\begin{equation}
\delta v(r) \sim U_0 F(r)^{1/3}G(r)^{h-1/3}
\label{6.31}
\end{equation}
obtained by the two definitions   (\ref{eq:33.a})  and (\ref{eq:monstre}).
In the equation (\ref{6.31}) $h$ is now the standard multifractal
scale-independent exponent.
Generalized scaling (\ref{6.31}) should be considered realized with
probability
\begin{equation}
P_h(r)\sim G(r)^{3-D(h)}
\label{6.32}
\end{equation}
inserting (\ref{6.31}) into (\ref{6.27}) we obtain:
\begin{equation}
r_d U_0 F(r_d)^{1/3} G(r_d)^{h-1/3} \sim \nu
\label{6.33}
\end{equation}
where $ r_d(h)$ is the fluctuating cutoff.\\
We now look for a solution of equation (\ref{6.33}) in a region where
\begin{equation}
F(r)\sim \frac{r^3}{L\eta_k ^2}
\label{6.34}
\end{equation}
\begin{equation}
G(r)\sim \frac{\eta_k}{L}
\label{6.35}
\end{equation}
After some algebra we obtain
\begin{equation}
r_d(h) \sim Re^{\frac{3h-1}{8}}\eta_k
\label{6.36}
\end{equation}
Thus $ r_d(h)$ is a fluctuating quantity as in (\ref{6.29}). These
fluctuations, however, happen in the region where $ \delta v(r)\sim r$ and
therefore no effect on the scaling of structure function is produced. From
(\ref{6.31}-\ref{6.34}) we can compute the scaling of $ \langle \Gamma^p
\rangle$ as function of Re. The scaling is independent on $ r_d(h)$ and it is:
\begin{equation}
\langle \Gamma^p\rangle \sim Re^{{3 \over 4}(p-\zeta_p)}
\label{6.37}
\end{equation}
consistent with (\ref{6.31}) in the limit $ r\rightarrow 0$.\\
Note that $ \langle \Gamma^2\rangle\sim Re^{{3 \over 4}
(2-\zeta_2)}$. This implies that
for $ \zeta_2 \ne 2/3$, $ \langle \Gamma^2\rangle$ does not scale as  Re.
If we want to recover the experimental fact that $ \langle \epsilon\rangle$ is
constant with Re, we should allow for a Re-dependent constant in
(\ref{6.31}).\\
At any rate, because $ \zeta_2-2/3$ is a small quantity, these effects are
quite small in the full range of available Re-number.\\
We can summarize the scenario I as follows: the scaling (\ref{6.25}) and
(\ref{6.31}) are verified to all scales; the condition (\ref{6.27}) introduces
a viscous cutoff which fluctuates in the region where $ \delta v(r)\sim r$;
intermittency in the gradient of the velocity field are prescribed by
(\ref{6.31}).\\
The above conclusions imply that scaling (\ref{6.22}) must be violated near
the viscous cutoff, as one can immediately check by an explicit computation.
One can take an opposite point of view and assume that (\ref{6.22}) is a
fundamental relationship which must not be violated. This corresponds to the
second scenario.

In the second scenario (\ref{6.27}) is disregarded and one generalize
(\ref{6.28}) as:
\begin{equation}
\nu \Gamma^2 \sim \frac{\delta v^3(r_d)}{\langle \delta v^3(r_d)
\rangle}\epsilon
\label{6.38}
\end{equation}
where $ \Gamma$ is the velocity gradient and $ r_d$ is the viscous cutoff.
In order to compute $ r_d$, one observes that $ \Gamma\sim\frac{1}{\tau (r_d)
}$ where $ \tau (r_d)$ is the eddy turnover time at the viscous cutoff.
We obtain:
\begin{equation}
\nu \sim \frac{\delta v^3(r_d)\tau (r_d)^2}{\langle \delta v^3(r_d)
\rangle}\sim\frac{\delta v(r_d)\langle \delta v^3(r_d)\rangle}{\epsilon}
\label{6.39}
\end{equation}
where following (\ref{6.23}), we used $ \tau (r_d)\delta v(r_d)=\langle\delta
 v^3(r_d)\rangle/ \epsilon$. Once again, by using (\ref{6.31}), (\ref{6.34})
and (\ref{6.35}) we can obtain an explicit formula for $ r_d(h)$:
\begin{equation}
r_d(h)\sim Re^{-\frac{3}{16}(\frac{13}{3}-h)}\eta_k
\label{6.40}
\end{equation}
Thus also in scenario II, we have strong fluctuations of the viscous
cutoff.\\
The computation of the gradients is quite straightforward from (\ref{6.38}).
We have:
\begin{equation}
\Gamma\sim Re^{1/2}\left(\frac{\delta v^3(r)}{\langle \delta v^3(r)\rangle}
\right)^{1/2}\sim Re^{1/2} G(r_d)^{\frac{3h-1}{2}}
\label{6.41}
\end{equation}
where we have used (\ref{6.31}). Finally by using (\ref{6.32}) and (\ref{6.35})
we get:
\begin{equation}
 \langle \Gamma^p\rangle = Re^{p/2} G(r_d)^{(\zeta_{3p/2} -p/2)}\simeq Re
 ^{p/2}Re^{3/4(\zeta_{3p/2} -p/2)}
\label{6.42}
\end{equation}
Note that in scenario II $ \langle \Gamma^2\rangle\sim Re$ because
$ \zeta_3=1$.\\
Because of (\ref{6.42}), the II scenario violates (\ref{6.31}) and (\ref{6.25})
for scales smaller than $ r_d(h)$ while (\ref{6.22}) is always satisfied.

It is quite difficult to understand which one of the two scenario is actually
verified by experimental and numerical data. In most cases the scale
resolution does not reach the region where $ \delta v(r)\sim r$. At any rate,
either (\ref{6.22}) or (\ref{6.25}) should be violated at very small scales as
the result of viscous effects. This violation is rather small and may not be
easily detectable at low or moderate Reynolds numbers.\\
The common point about the two scenario is that the viscous cutoff (if any)
acts at scale where already the velocity structure functions behave in a
regular way, i.e. $\delta v(r)\sim r$.\\
\bigskip

\section{Conclusions}

In this paper we have proposed several new results concerning
the scaling behaviour of small scale statistical properties of
turbulence. It is worth to summarize our main findings trying
to outline questions which are still to be answered.

1) We have reviewed the main results on ESS and in particular we have
shown that in homogeneous and isotropic flows in turbulence, Rayleigh
Benard convection and solar wind magnetohydrodynamics, the ratio
$\zeta(p) / \zeta(3)$ seems to have an universal behaviour. This is
a rather striking and unexpected result which implies that anomalous
violation of dimensional scaling may be explained in an universal way.
We do not know any simple phenomenological explanation for our finding.

2) We have shown that ESS is not observed when
relatively strong shear flows are present. A phenomenological analysis,
based on the Kolmogorov equation, shows the relevance of a length scale
based on the mean energy dissipation and the shear strength. This analysis
should be refined in order to acquire more quantitative predictions. At
any rate, our observation suggests that  previous finding of violations
of ESS should be due to the presence of shear flows.

3) We have shown that the refined Kolmogorov similarity can be generalized
by including ESS. This generalization is verified extremely well in both
experiments and numerical simulations. More important, we have shown that
the generalized refined Kolmogorov similarity is true also in cases where
ESS is not observed.

4) Similar to the previous point, we have shown that the hierarchy
relation based on log-poisson distribution for the structure functions
is very well supported by experimental data, also for very small scales
where ESS is not observed.

5) Based on our results in 1)-4) we have proposed a generalization of
ESS. This generalization is supported both by experimental and
numerical data and it seems not affected by viscous cutoff.

6) We have developed a theory which unifies the previous point. The theory
is based on the assumption that the probability distribution is infinitively
divisible and predicts the existence of the generalized ESS. The theory
can also be used to generate artificial signals which displays all the
scaling features observed in real data.

7) We have shown that the original proposal on the multiscaling
for the viscous cutoff is incompatible with the turbulence data. The
theory formulated in this paper removes this incompatibility and
suggests that multiscaling is acting at much smaller scales than
previously proposed. The new point on the theory is a change of view
in the probability distribution of the original multifractal model
which is not directly linked to a geometrical interpretation in terms
of fractal dimensions.

8) Finally we have shown that violations of either the generalized
refined Kolmogorov similarity or the generalized ESS should occur
at very small scale. Our present data analysis does not allow us
to distinguish among the two possibilities.

\section{Acknowledgements}

We thank B. Castaign, S. Fauve, U. Frisch, Z.S. She,
K.R. Sreenivasan, G. Stolovitzky, ,
 for many useful discussions. During the writing of this
paper, we became aware of a manuscript of C. Meneveau dealing
with a possible explanation of ESS. His conclusions are physically
close to part of the discussion in section 6.

This work has been partially supported by EEC contracts ERBCHBICT941034,
CT93-EVSV-0259 and CHRX-CT94-0546.

\newpage

{\bf Figure captions}

\bigskip

{\bf Figure 1: } Structure functions  $S_p(r)/r^{\zeta(p)}$ as a
function of $r$.  Data taken from an experiment on a jet
at $R_\lambda =800 $ (diamonds). Data  taken from the wake behind
a cylinder at $R_\lambda =140 \ $ (triangles).
(a) p=6 and $\zeta(6)= 1.78$. (b) p=3 and  $\zeta(3)=1$. Logarithms
are in  base 10.

{\bf Figure 2: }  Structure functions $S_6$ as a function  of
$S_3$ at $R_\lambda=800$ (a)   and $R_\lambda=140 $ (b)
computed from the same data-set of fig.1. Vertical dashed
lines indicate the value of $S_3$ at $5 \eta_k $.

{\bf Figure 3: } Logarithm  of ratio of the
universal functions $f_n/f_2$ for two cases $n=6$ (diamonds)
and $n=4$ (circles)
for the wake behind the cylinder  of fig.1
 as a function of $r/\eta_k$.

{\bf Figure 4: }  Dependence of the exponent $\beta(6,3)$
as a function of $ Re$. ($R_\lambda \simeq 1.4 Re^{1/2}$).
The last point is from ref. \cite{Anselmet}. See also ref. \cite{Tab1995}.

{\bf Figure 5: }  Log-log plot of $<\epsilon_r^2> \ S_3(r)^2$
against $S_6(r)$ at $R_\lambda=500$. The straight line refers to
the slope 1.005. Data are from an experiment of turbulence
behind a cylinder and the measurement point was at about 25 diameter
down stream.

{\bf Figure 6: }
a) Log-log plot of ESS scaling for the longitudinal structure function
$S_6(r)$ versus $S_3(r)$. Data are taken from a numerical simulation of
a shear flow at $R_\lambda=40$. The dashed
line is the best fit with slope $ 1.79$.  Every point in the plot corresponds
to a grid point and the lattice spacing is $ \sim 1 \eta_k$ wide.
The computation of the structure functions
is performed in points of the flow where the shear has a minimum.

b) the same of (a) but for points where the shear is maximum. The dashed
line is the best fit with slope $ 1.43$.
At variance with previous case ESS is not observed.

{\bf Figure 7: } Check of eq.(\ref{2.3}) for p=6 using the same numerical
simulation of the shear flow  discussed in fig.6 (log-log plot).
Energy dissipation has been computed by using the 1-dimensional surrogate
in order to compare this result with laboratory experiments (see fig.5).
(a) points of minimum shear.
(b)  points of maximum shear.
The points refer to the scales at $2,4,5,8,10,16,20,32,
 40$ grid points and the dashed line is the best fit done over these points,
 corresponding to the slope $ 0.99$ for both minimum and maximum shear data.
Although in this case ESS is not observed
(see fig. 6b), the generalized refined Kolmogorov hypothesis eq.(\ref{2.3})
works within $3\%$.

 {\bf Figure 8:} The function $F_{p+1}(r)$ defined in eq.(\ref{4.3}) is
plotted, for several values of $p$,
as a function of  $ \big[ \big( { F_{p}(r) }
\big) ^{\beta '}
\cdot  \tilde F(r) \big] \ $ with $\beta = 2/3$,
$\beta '= \beta^\delta  \ $
and $\delta=1/3$ .
 $R_\lambda=140 \  $ in (a) and  $R_\lambda=800$ in (b).
The $5$ curves in (a) and (b)  correspond to
$p=1,2,3,4,5$ starting from the bottom lines. They have been
 vertically shifted
 of $-0.4,-0.2,0,0.2,0.4$ in order to separate them.
 The solid lines have
slope 1. Logarithms are base 10.

{\bf Figure 9: } Log-Log plot of $G_6(r)$ versus $G_5(r)$ for different
laboratory and numerical experiments. ($+$)
Data are taken in a wake behind a
cylinder where standard  ESS was not observed \cite{Benzi 1993B}.
($\circ $)
Data taken from the region with log-profile of a boundary layer
(courtesy of G. Ruiz Chavarria) where standard ESS was not observed.
(squares) Data taken from  a numerical simulation of thermal convection
\cite{Benzi 1994B} where standard ESS was observed.( $\Delta$ ) Data
taken from a direct numerical simulation of a channel flow where
standard ESS was not observed\cite{Benzi 1995B}.

{\bf Figure 10:}
Log-log
plot of $F_{p+1}$ against $F_p$ for $p=1,...6$ and
$r= 3 \eta_k $ (squares) and $r= 30 \eta_k$ (circles) for
the case of two numerical simulations, namely Rayleigh Bernard thermal
convection (a) and channel flow (b). As one can see a clear scaling is
observed with the same scaling exponent $\beta '$ both for small and
relatively large values of $r$. This confirms the quality of the
generalized ESS scaling (\ref{5.3}).

{\bf Figure 11:} Log-log plot  of the 6th order structure function
for the signal A with 19 fragmentation at small
Reynolds number. Notice the absence of any scaling range.

{\bf Figure 12:} G-ESS ($\log(G_6(r))$ vs $\log(G_4(r))$)
for the signal A at the same Reynolds number
of figure 10. The slope is $\rho_{4,6}$ = 0.241 in perfect
agreement with the theoretical prediction obtained from (\ref{5.4})
where for $\zeta(p)$ we have used the She-Leveque expression.

{\bf Figure 13:} 6th order local scaling exponents for the signal A (circles)
and signal B (squares).
 Notice that the qualitative behaviour of the two signals
is almost the same: both of them go from an intermittent
scaling ($\zeta(6) \sim 1.8$) at large scale to a laminar scaling
($\zeta(6)=6$) at small scales.

{\bf Figure 14:} Generalized-Kolmogorov refined hypothesis
(\ref{6.22})  for the 6th order
structure function in both signal A (squares) and signal B (circles).
Notice the sudden jump at the Kolmogorov scale present when  multiscaling
is valid (signal B).

{\bf Figure 15:} Compensated slope ($G_4(r)/G_6^{\rho_{4,6}}(r)$) for
signal B. Notice deviations of the order of $10 \%$ while for signal A
the same quantity is  constant by definition.

\vfill
\eject
\centerline{\bf Table I}
 \begin{table}[htp]
 \begin{center}
 \begin{tabular}{|r|r|r|r|r|r|r|}
 \hline
 \hline
       & p
       & $ \zeta _p$
       & $ \zeta _p$(Bolg.)
       & $ \zeta _p$(MHD)
       & $ \beta (p,3)$(Bolg)
       & $ \beta (p,3)$(MHD)
 \\
 \hline
       & $1$
       & $0.37$
       & $0.77$
       & $0.28$
       & $0.37$
       & $0.36$
 \\
\hline
       & $2 $
       & $ 0.70$
       & $1.46$
       & $0.55$
       & $0.70$
       & $0.70$
 \\
\hline
       & $ 3$
       & $ 1$
       & $2.08$
       & $0.78$
       & $1$
       & $1$
 \\
\hline
       & $ 4$
       & $ 1.28$
       & $2.66$
       & $1$
       & $1.28$
       & $1.28$
 \\
\hline
       & $ 5$
       & $ 1.54$
       & $3.20$
       & $1.20$
       & $1.54$
       & $1.54$
 \\
\hline
       & $ 6$
       & $ 1.78$
       & $3.70$
       & $1.39$
       & $1.78$
       & $1.78$
 \\
\hline
       & $ 7$
       & $ 2.00$
       & $4.16$
       & $1.58$
       & $2.02$
       & $2.02$
 \\
\hline
       & $ 8$
       & $ 2.23$
       & $4.63$
       & $1.75$
       & $2.24$
       & $2.24$
 \\
\hline
\hline
\end{tabular}
\caption[]{We show some measured values of $ \zeta(p)$ and $ \beta(p,3)$
for $ 1\leq p\leq 8$.
In the second column we report the $\zeta(p)$ measured in
3D homogeneous and isotropic turbulence($30<R_\lambda<2000$),
in the third
column the $\zeta(p)$ measured in Rayleigh Benard convection when the
Bolgiano scaling is the relevant one($R_\lambda\simeq30$),
in  the  fourth column the $\zeta(p)$ obtained from the measurements of the
solar wind ($R_\lambda\simeq 5 \ 10^6$).
We note that the $\zeta(3)$ of the last two cases  are clearly very
different from $1$ which is the value of $\zeta(3)$  in the second column.
The ratios $\beta(p,3)$ computed from the values of the third and fourth
column are shown in the fifth and sixth columns respectively.
The $\beta(p,3)$ are equal within error bars to those of the first column. }
 \end{center}
 \label{tab}
 \end{table}

 \end{document}